\RequirePackage{silence}
\documentclass[twocolumn,fleqn,usenatbib,useAMS]{aastex631}
\acceptjournal{ApJ}
\usepackage{graphicx}
\usepackage{amsmath}

\usepackage{amsfonts}
\usepackage{float}
\usepackage{bm}
\usepackage{ae,aecompl}
\usepackage{array}
\usepackage{soul}
\usepackage{mathtools}
\usepackage{multirow}
\usepackage[utf8]{inputenc}
\usepackage{booktabs}
\usepackage{graphicx}

\shorttitle{El Gordo challenges $\Lambda$CDM for any plausible velocity}
\shortauthors{E. Asencio, I. Banik, and P. Kroupa}

\begin{document}

\title{The El Gordo Galaxy Cluster Challenges $\Lambda$CDM for Any Plausible Collision Velocity}


\author[0000-0002-3951-8718]{Elena Asencio}
\affiliation{Helmholtz-Institut f\"ur Strahlen und Kernphysik (HISKP), University of Bonn, Nussallee 14$-$16, D-53115 Bonn, Germany}
\author[0000-0002-4123-7325]{Indranil Banik}
\affiliation{Scottish Universities Physics Alliance, University of Saint Andrews, North Haugh, Saint Andrews, Fife, KY16 9SS, UK}
\author[0000-0002-7301-3377]{Pavel Kroupa}
\affiliation{Helmholtz-Institut f\"ur Strahlen und Kernphysik (HISKP), University of Bonn, Nussallee 14$-$16, D-53115 Bonn, Germany}
\affiliation{Astronomical Institute, Faculty of Mathematics and Physics, Charles University, V Hole\v{s}ovi\v{c}k\'ach 2, CZ-180 00 Praha 8, Czech Republic}
\correspondingauthor{Elena Asencio}
\email{s6elena@uni-bonn.de}



\begin{abstract}

El Gordo (ACT-CL J0102-4915) is an extraordinarily large and bright galaxy cluster collision. In a previous study, we found that El Gordo is in $6.2\sigma$ tension with the $\Lambda$ cold dark matter ($\Lambda$CDM) standard model when assuming the nominal mass and infall velocity values from the hydrodynamical simulations of Zhang et al. ($M_{200} = 3.2 \times 10^{15} M_{\odot}$ and $V_{\textrm{infall}} = 2500~\textrm{km~s}^{-1}$, respectively). The recent weak-lensing study of Kim et al. showed that the mass of El Gordo is actually $2.13^{+0.25}_{-0.23} \times 10^{15} M_{\odot}$. Here we explore the level of tension between El Gordo and $\Lambda$CDM for the new mass estimate, assuming several $V_{\textrm{infall}}$ values. We find that in order to reduce the tension below the $5\sigma$ level, the El Gordo subclusters should have $V_{\textrm{infall}} < 2300~\textrm{km~s}^{-1}$ ($V_{\textrm{infall}} < 1800~\textrm{km~s}^{-1}$ when considering the combined tension with the Bullet Cluster). To the best of our knowledge, the El Gordo hydrodynamical simulations conducted so far require $V_{\textrm{infall}} \geq 2500~\textrm{km~s}^{-1}$ to simultaneously reproduce its morphology and its high X-ray luminosity and temperature. We therefore conclude that El Gordo still poses a significant challenge to $\Lambda$CDM cosmology. Whether the properties of El Gordo can be reconciled with a lower $V_{\textrm{infall}}$ should be tested with new hydrodynamical simulations that explore different configurations of the interaction.

\end{abstract}


\keywords{galaxies: clusters: individual (El Gordo) -- galaxies: clusters: individual (Bullet Cluster) -- large-scale structure of universe -- gravitation -- dark matter -- methods: statistical}

\section{Introduction}
\label{Introduction}

El Gordo (ACT-CL J0102-4915) is a supermassive interacting galaxy cluster composed of two infalling subclusters observed at redshift $z=0.87$. It was found by the Atacama Cosmology Telescope (ACT) collaboration in a 455 deg$^2$ survey \citep{Menanteau_2010, Marriage_2011}. El Gordo is well known for its extreme properties of mass, redshift, and infall velocity, which are remarkably rare in the context of the $\Lambda$ cold dark matter ($\Lambda$CDM) standard model of cosmology \citep{Menanteau_2012, Katz_2013, Jee_2014}.

The properties of El Gordo were constrained for the first time by \citet{Menanteau_2012} from the analysis of additional multi-wavelength observations. Those authors described El Gordo as the most massive, hottest, most X-ray luminous, and brightest Sunyaev-Zel’dovich (SZ) effect cluster known at redshift $z > 0.6$. They also noted the presence of a wake formed in the X-ray surface brightness caused by the passage of one subcluster through the other. Their estimated values for the redshift, mass, and relative velocity between the two subclusters at the time of observation are: $z = 0.87$, $M_{200a} = 2.16 \pm 0.32 \times 10^{15} h^{-1}_{70} M_{\odot}$, and $V_{\textrm{obs}} = 1200 - 2300~\textrm{km~s}^{-1}$, respectively, where $h_{70}$ is the Hubble constant $H_0$ in units of 70~\textrm{km~s}$^{-1}$~Mpc$^{-1}$ and $M_{200a}$ is the mass within the radius such that the average enclosed density is $200\times$ the cosmic average, which at that epoch was slightly below the critical density. The estimated range of $V_{\textrm{obs}}$ was obtained from the line-of-sight peculiar velocity difference between the two subclusters ($\Delta V_{\textrm{pec}} = 586 \pm 96~\textrm{km~s}^{-1}$) and from assuming that the collision is likely taking place close to the plane of the sky (inclination angle $\theta = 15^\circ - 30^\circ$), leading to $V_{\textrm{obs}} = \Delta V_{\textrm{pec}}/\sin \theta$. \citet{Menanteau_2012} justified the low adopted $\theta$ by noting that we can only see the El Gordo morphological features so prominently if the interaction is nearly in the plane of the sky. The low $\theta$ assumption is also supported by the high degree of polarization in the radio relics of El Gordo \citep{Lindner_2014}. \citet{Menanteau_2012} assumed that the infall velocity of the two subclusters ($V_{\textrm{infall}}$) should be about the same as $V_{\textrm{obs}}$. However, the present velocity is not necessarily a good guide to the pre-merger velocity due to dynamical friction and gas drag, among other processes. $N$-body/hydrodynamical simulations are needed to get a better constraint on $V_{\textrm{infall}}$.

\citet{Karen_2015} performed a series of $N$-body simulations of El Gordo that focused on estimating its particular dynamical and kinematic features. They aimed to constrain $\theta$, $V_{\textrm{obs}}$, and the pericenter velocity $V_{\textrm{peri}}$, among other parameters. Their estimated $\Delta V_{\textrm{pec}} = 476 \pm 242~\textrm{km~s}^{-1}$ is lower than that observed by \citet{Menanteau_2012} but consistent within uncertainties. From the polarization fraction of El Gordo's synchrotron radiation \citep{Lindner_2014}, \citet{Karen_2015} obtained an upper limit to the inclination angle of $\theta \le 35^\circ$. Taking as inputs a mass of $M_{200} = 2.06 \times 10^{15}$ $M_{\odot}$, a mass ratio of 1.71, a projected separation of $0.740 \pm 0.007$~Mpc between the subclusters, and their estimated value for $\Delta V_{\textrm{pec}}$, their best-fit model obtained with the Markov Chain Monte Carlo (MCMC) method yields $\theta = 21^{\circ} \pm ^{9}_{11}$, $V_{\textrm{obs}} = 940^{+860}_{-580}~\textrm{km~s}^{-1}$, and $V_{\textrm{peri}} = 2400_{-200}^{+400}~\textrm{km~s}^{-1}$. In their model, the value of $V_{\textrm{infall}}$ at a subcluster separation equal to the sum of the two virial radii roughly corresponds to $50\%$ of the $V_{\textrm{peri}}$ value \citep[see section 5.1 of][]{Karen_2015}, which means that they would have obtained $V_{\textrm{infall}} \approx 1200~\textrm{km~s}^{-1}$ for their best-fit $V_{\textrm{peri}}$ value. It is worth noting that in their MCMC analysis, they exclude all realizations that would result in $V_{\textrm{peri}}$ higher than the escape velocity. Unlike other studies of El Gordo, \citet{Karen_2015} found that the best-fitting configuration is that in which the two subclusters already had a pericenter passage and are returning toward a second passage.

Hydrodynamical simulations of El Gordo aim to understand the detailed gas dynamics required to reproduce the SZ and X-ray observables of this cluster, including its twin-tailed morphology. The first hydrodynamical simulation of El Gordo was carried out by \citet{Donnert_2014}. Using the results of \citet{Menanteau_2012}, they chose as initial parameters: total mass $M_{200} = 2.71 \times 10^{15} M_{\odot}$, mass ratio of 2.35, main cluster radius $R_{200,1} = 2.550$~Mpc, secondary cluster radius $R_{200,2} = 1.925$~Mpc, $V_{\textrm{infall}} = 2600~\textrm{km~s}^{-1}$ at a distance of 5.23~Mpc, and a small impact parameter of $b = 20$~kpc. Though they were not able to reproduce the X-ray wake in their simulation, they managed to match the observed X-ray luminosity of $L_X = 2.19 \pm 0.11 \times 10^{45}~h^{-2}_{70}$ erg~s$^{-1}$ \citep{Menanteau_2012} very accurately when the observed dark matter core separation is $d_{\textrm{core}} = 690$~kpc, similar to the observed separation of $\approx 700$~kpc \citep{Jee_2014}.

The hydrodynamical simulations of \citet{Molnar_2015} used a total mass of $M_{200} = 2.15 \times 10^{15}$ $M_{\odot}$ with mass ratio 1.87 and radii of $R_{200,1} = 2.304$~Mpc and $R_{200,2} = 1.944$~Mpc for the individual subclusters. Those authors ran several simulations for different $V_{\textrm{infall}}$ (at a distance of 4.25~Mpc) and $b$ in order to find the best match to observations of El Gordo. They found that for $V_{\textrm{infall}} = 2250 \pm 250~\textrm{km~s}^{-1}$ and $b = 300^{+50}_{-100}$~kpc, they could reproduce the twin-tailed morphology of El Gordo, its relative radial velocity, and the relative positions of its X-ray, SZ, and weak-lensing peaks. However, they could not match the high observed X-ray luminosity and temperature of El Gordo \citep{Zhang_2015}. Moreover, their best-fit model assumes an inclination angle of $\theta = 45^\circ$, which is well above the maximum value of $\theta = 35^\circ$ as constrained by \citet{Karen_2015} from the polarization fraction. However, they obtained $V_{\textrm{obs}} = 1009~\textrm{km~s}^{-1}$ for this model, very consistent with the value obtained by \citet{Karen_2015}. With this, \citet{Molnar_2015} also showed that the $V_{\textrm{infall}} \approx V_{\textrm{obs}}$ assumption made by \citet{Menanteau_2012} clearly underestimates the value of $V_{\textrm{infall}}$: according to their simulations, $V_{\textrm{infall}} \approx 1.5 \, V_{\textrm{obs}}$.

The latest and most complete exploration of the parameter space using hydrodynamical simulations of El Gordo was done by \citet{Zhang_2015}. These are also the simulations that better manage to reproduce the observed properties of El Gordo as they simultaneously reproduce its morphology, luminosity, temperature, and relative radial velocity. \citet{Zhang_2015} test two different scenarios for the interaction. The first scenario (referred to as Model A by the authors) assumes that the interaction is a very energetic head-on collision between the two subclusters. \citet{Donnert_2014} and \citet{Molnar_2015} also assume this scenario in their simulations. The best-fit parameters found by \citet{Zhang_2015} in their Model A are: $M_{200} = 1.95 \times 10^{15} M_{\odot}$, mass ratio of 2, $R_{200,1} = 1.66$~Mpc, $R_{200,2} = 1.32$~Mpc, $V_{\textrm{infall}} = 3000~\textrm{km~s}^{-1}$ (at 5.84~Mpc), and $b = 300$~kpc, viewed at an inclination angle $\theta = 75^\circ$. With these parameters, they could generate X-ray surface brightness and mass density distributions that match the observations. But the twin-tailed X-ray morphology that they obtained is smaller and more asymmetric than observed and only appears when the projected distance between the clusters is $\approx 600$~kpc. Also, the luminosity for this model ($L_X = 2.48 \pm 0.03 \times 10^{45}~h^{-2}_{70}$ erg~s$^{-1}$) is a bit higher than observed, while the inclination angle surpasses significantly the maximum estimated value \citep{Karen_2015}.

In their second scenario (Model B), \citet{Zhang_2015} assume that the interaction took place as a less violent, off-centered collision. The best-fit parameters in this Model B case are: $M_{200} = 3.19 \times 10^{15} M_{\odot}$, mass ratio of 3.6, $R_{200,1} = 2.06$~Mpc, $R_{200,2} = 1.35$~Mpc, $V_{\textrm{infall}} = 2500~\textrm{km~s}^{-1}$ (at 6.72~Mpc), and $b = 800$~kpc, viewed at an inclination angle of $\theta = 30^\circ$. With these parameters, the model reproduces the temperature and X-ray luminosity of El Gordo and a two-tailed X-ray morphology bearing closer resemblance to the observed morphology, especially concerning the size and symmetry of the tails and the projected separation of the clusters ($\approx 780$~kpc). Since this model performs better than fiducial Model A at reproducing the morphological properties of El Gordo, \citet{Zhang_2015} chose this set of parameters as their preferred model.

In our previous work \citep*{Asencio_2021}, we calculated the probability of finding a pair of clusters in $\Lambda$CDM cosmology at $z = 1$ with the mass, mass ratio, and infall velocity given by the best-fit pre-merger parameters of Model B \citep{Zhang_2015} in the survey region in which the El Gordo properties were first constrained \citep[sky area of 755 deg$^2$;][]{Menanteau_2012}. We found that the probability of such a pre-merger configuration arising in $\Lambda$CDM is only $7.51 \times 10^{-10}$, which corresponds to a $6.16\sigma$ failure for a 1D Gaussian. In other words, according to our statistical analysis, only one El Gordo analog is expected in a volume of $\approx 5.6 \times 10^8~\textrm{cGpc}^3$. This is $\approx 10^9 \times$ larger than the effective (from $z=1$ to $z=1.06$) survey volume of El Gordo, and $\approx 5 \times 10^7 \times$ larger than the effective full-sky volume. The probability is even lower if we take into account that El Gordo is not the only object whose properties pose a problem for the $\Lambda$CDM model. The Bullet Cluster is another interacting galaxy cluster known for its very high infall velocity \citep[$\approx 3000~\textrm{km~s}^{-1}$;][]{Lage_Farrar_2014}. From \citet{Kraljic_2015}, we inferred that the probability of observing the Bullet Cluster in the survey region where it was discovered is $5.4 \times 10^{-3}$ ($2.78\sigma$). Combining the probability of observing both El Gordo and the Bullet Cluster in their respective survey regions, we obtained a total probability of $1.24 \times 10^{-10}$ ($6.43\sigma$). We also discussed the impact that the different uncertainties in the analysis assumptions could have had on our nominal result. These caused the final result to differ somewhat from the nominal $6.43\sigma$ value. But for the most plausible assumptions, we always obtained a tension $>5\sigma$. From these results, we concluded that the $\Lambda$CDM model is falsified at high significance ($>5\sigma$) $-$ assuming the best-fit parameters of \citet{Zhang_2015} $-$ whether we consider El Gordo alone or together with other problematic clusters.

However, a very recent and detailed weak-lensing study of El Gordo \citep{Kim_2021} obtained a lower estimate for its mass and mass ratio. In Section \ref{Analysis}, we obtain the tension between El Gordo and $\Lambda$CDM for the new mass and mass ratio values, assuming several different infall velocities. In Section \ref{Discussion}, we discuss our results and their implications for $\Lambda$CDM in light of the $V_{\textrm{infall}}$ values obtained in prior numerical and hydrodynamical simulations of El Gordo. We conclude in Section \ref{Conclusions}.

\section{Analysis and results}
\label{Analysis}

The weak-lensing study of \citet{Kim_2021} found that El Gordo has a mass of $M_{200} = 2.13^{+0.25}_{-0.23} \times 10^{15} \, M_{\odot}$ and a mass ratio of 1.52. The mass estimate of $M_{200} \approx 2 \times 10^{15} \, M_{\odot}$ is supported by recent strong-lensing analyses \citep{Caminha_2023, Diego_2023}. \citet{Kim_2021} claim that using the new mass estimate and assuming $V_{\textrm{infall}} \approx 450~\textrm{km~s}^{-1}$ at a separation of 5.5~Mpc, the tension with the $\Lambda$CDM model disappears. This is technically true since repeating the analysis performed in \citet{Asencio_2021} for the nominal values of mass, mass ratio, and infall velocity assumed in \citet{Kim_2021} increases the probability of finding an El Gordo-like object in $\Lambda$CDM to $6.26 \times 10^{-3}$ (2.73$\sigma$).

However, the significant decrease of the tension with respect to our previous study \citep{Asencio_2021} is mainly caused by the lower infall velocity assumption and not by the lower mass estimate. \citet{Kim_2021} obtained this $V_{\textrm{infall}}$ by performing a similar analysis to that of \citet{Karen_2015}, but also accounting for some of the inaccuracies present in this analysis. The main problems that they identified in \citet{Karen_2015} were the assumption that dynamical friction is negligible in the El Gordo case, an incorrect sampling of the angle $\theta$ in their MCMC analysis, and the fact that they allow for merger scenarios in which the two subclusters start to free-fall at a distance similar to the present subcluster separation. In their corrected version of the analysis, they found that the scenario in which El Gordo's subclusters are returning for a second pericenter passage is no longer favored. As in \citet{Karen_2015}, they also impose an upper limit to their MCMC analysis that rules out all models with collision velocities higher than their escape velocity \citep[$\approx 3600~\textrm{km~s}^{-1}$ at the current subcluster separation of 0.7~Mpc for the parameters found by][]{Kim_2021}\footnote{This corresponds to an escape velocity of $\approx 1200~\textrm{km~s}^{-1}$ at $6.06$~Mpc, the distance at which we obtain $V_{\textrm{infall}}$ in our analysis}, which is the main reason they obtain such a low $V_{\textrm{infall}}$. This constraint is motivated by the fact that previous studies from cosmological simulations had already shown that it is highly unlikely that the pairwise velocity of massive merging clusters ($M \ga 10^{15} \, M_{\odot}$) exceeds 3000~km~s$^{-1}$ in a $\Lambda$CDM cosmology \citep{Lee_Komatsu_2010, Thomson_Nagamine_2012, Karen_2015}. Since the analysis of \citet{Kim_2021} imposes as a prior that we should not consider any $V_{\textrm{infall}}$ value which is unlikely to arise in $\Lambda$CDM, the $V_{\textrm{infall}}$ value they obtain cannot be used to estimate the likelihood of observing El Gordo in $\Lambda$CDM.



As it is not possible to infer the actual infall velocity needed to reproduce the observed morphology of El Gordo for the new mass estimate without performing a more detailed study using $N$-body or hydrodynamical simulations, we obtain the level of tension between $\Lambda$CDM and El Gordo assuming $M_{200} = 2.13^{+0.25}_{-0.23} \times 10^{15} \, M_{\odot}$, a mass ratio of 1.52, and several different values of $V_{\textrm{infall}}$ at a distance of 6.06~Mpc using the nominal `lightcone tomography' technique we presented in \citet{Asencio_2021}. The method can be briefly summarized as follows: the probability of observing an El Gordo-like object in the survey region according to $\Lambda$CDM is determined by looking for pre-merger El Gordo analogs in a large $N$-body cosmological simulation. The conditions required for a cluster pair to be a candidate are: (i) mass ratio below that of El Gordo (1.52 for the model that we test in this study); and (ii) infall velocity/escape velocity ratio above that of El Gordo at twice the sum of its virial radii. This lets us obtain $\log_{10} N \left( \geq \log_{10} M/M_{\odot} \right)$, the number of candidate pairs above a certain mass (in logarithmic scale). This function is very well fitted by a parabola. From this fit, the frequency of very high-mass candidates $-$ for which there are no analogs in the simulation box $-$ can be extrapolated. This process is repeated at three different redshifts. This allows us to infer the relation between the redshift and the number of analogs $N$ above a certain minimum mass. From this, one can obtain the number of El Gordo analogs in any mass and redshift range. We estimate that the pre-merger configuration took place at $z=1$ and take $M_{200} = 2.13 \times 10^{15} \, M_{\odot}$ for the El Gordo mass. The number of analogs then has to be scaled to account for the fact that the survey volume in which El Gordo was found is smaller than the simulation box volume. The number $N$ of analogs which have $M$ and cosmic scale factor $a \equiv 1/\left( 1 + z \right)$ in a less likely region of parameter space than El Gordo is then converted into a probability using Poisson statistics: $P = 1 - \exp \left( -N \right)$. This probability is related to the equivalent number of standard deviations for a normal distribution by the Gaussian equation: $1 - \frac{1}{\sqrt{2\mathrm{\pi}}} \int_{-\chi}^{\chi} \exp \left( -x^2/2 \right) dx \equiv P$.

\begin{figure}
	\centering
	\includegraphics[width = 8.5cm]{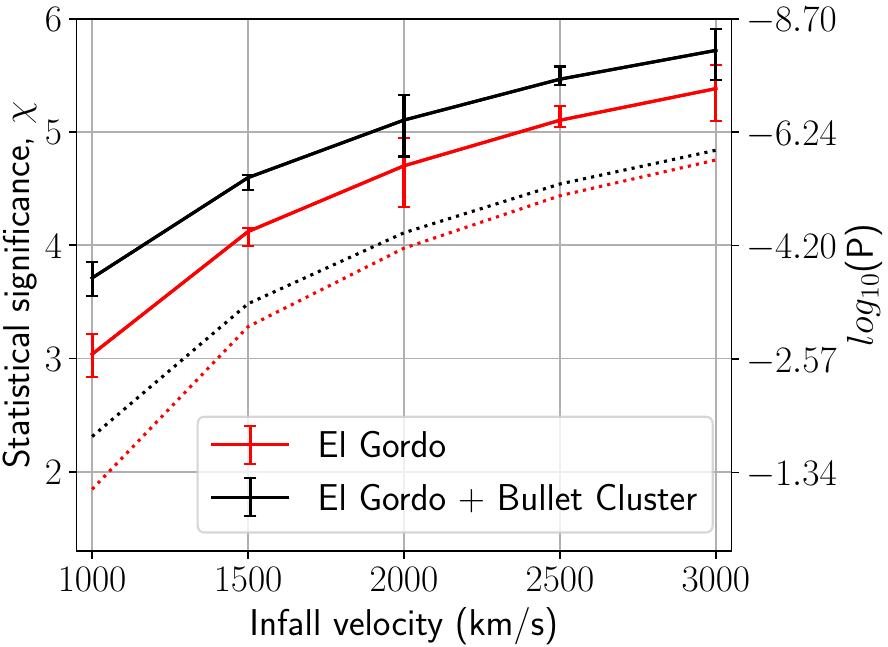}
	\caption{The equivalent number of standard deviations for a Gaussian ($\chi$) against the assumed El Gordo $V_{\textrm{infall}}$ at twice the sum of the virial radii for the El Gordo survey region (solid red line). For reference, the logarithmic probability $\log_{10} \left( P \right)$ corresponding to each $\chi$ is also shown on the right axis. The solid black line shows $\chi$ considering both El Gordo and the Bullet Cluster in their respective survey regions (with $\chi_{BC} = 2.78$). The error bars account for the uncertainty in the El Gordo mass \citep{Kim_2021}. The dotted lines in the same colors represent more conservative estimates assuming the respective survey regions cover the whole sky, in which case $\chi_{BC} = 1.65$ \citep{Kraljic_2015}.}
	\label{Results_plot}
\end{figure}

The results are shown in Figure~\ref{Results_plot}, where we have also plotted the level of tension for El Gordo and the Bullet Cluster combined. This can be approximately done by adding the square of the El Gordo $\chi$ value ($\chi_{EG}$) to $\chi_{BC}^2$, where $\chi_{BC} = 2.78$ is based on scaling the result of \citet{Kraljic_2015} to the sky area in which the Bullet Cluster was discovered \citep[see section~3.4 of][]{Asencio_2021}.

\section{Discussion}
\label{Discussion}

Figure~\ref{Results_plot} shows that the infall velocity required to get below $5\sigma$ in the El Gordo case is $\approx 2300~\textrm{km~s}^{-1}$. This falls to $\approx 1800~\textrm{km~s}^{-1}$ when considering both El Gordo and the Bullet Cluster. In the following, we discuss the implications of these results given the $V_{\textrm{infall}}$ values obtained in previous studies of El Gordo.

As mentioned in Section~\ref{Analysis}, the $V_{\textrm{infall}} \approx 450~\textrm{km~s}^{-1}$ value \citep{Kim_2021} is obtained for those models in which the collision velocity does not exceed the escape velocity. When testing for possible scenarios of the El Gordo interaction, there is no observational reason to assume that the two subclusters are gravitationally bound, as far as we are concerned. The argument that \citet{Kim_2021} provide to impose a limit to $V_{\textrm{infall}}$ is the ``timing argument'' \citep[also invoked by][]{Karen_2015}, the idea that structures in the Universe owe their peculiar velocities to interactions with other structures. Due to dynamical friction, when massive structures interact with less massive structures, kinetic energy in bulk motion is transformed into random motions. The maximum velocity that massive structures are expected to have can be inferred from cosmological simulations. These show that objects with $M > 10^{15} \, M_{\odot}$ and relative velocity $> 3000~\textrm{km~s}^{-1}$ are very unlikely to arise in the $\Lambda$CDM cosmology \citep{Lee_Komatsu_2010, Thomson_Nagamine_2012}. \citet{Kim_2021} and \citet{Karen_2015} imposed this upper limit to $V_{\textrm{infall}}$ since the escape velocity of El Gordo is of the same order as the maximum pairwise velocity found in $\Lambda$CDM cosmological simulations. This constitutes a circular argument, so we consider all studies that imposed this condition to be unsuitable for assessing the likelihood of observing El Gordo in the $\Lambda$CDM cosmology.

A collision velocity larger than the escape velocity can arise if the two subclusters are infalling on hyperbolic orbits and dynamical friction from the dark matter halos does not operate to slow down the relative velocity of the subclusters. It is also possible that the standard description of galaxy clusters as systems governed by Newtonian gravity and surrounded by a cold dark matter halo might be incorrect \citep[see section 4.3 of][]{Asencio_2021}. In that case, the infall velocity required to achieve a certain morphology could be different than in the standard model. Moreover, if the collisionless component surrounding the clusters was formed by sterile neutrinos (or some form of hot dark matter), the effect of dynamical friction would be less significant due to their higher velocity dispersion. Thus, if the $\Lambda$CDM description of galaxy clusters is incorrect, this could manifest as a particularly large $V_{\textrm{infall}}$ in $\Lambda$CDM models. Therefore, it is necessary to account for all $V_{\textrm{infall}}$ values in order to objectively assess the validity of $\Lambda$CDM with regards to the properties of the El Gordo cluster.




Moreover, $V_{\textrm{infall}} \approx 450~\textrm{km~s}^{-1}$ seems rather unlikely since all the El Gordo studies that used hydrodynamical simulations have shown that $V_{\textrm{infall}} \ga 2000~\textrm{km~s}^{-1}$ is required to reproduce its morphology and/or its strong X-ray emission \citep{Donnert_2014, Molnar_2015, Zhang_2015}. The X-ray emission in galaxy clusters is generally understood to be caused by thermal bremsstrahlung from hot gas \citep{Sarazin_1986}. In the case of El Gordo, the fact that the X-ray morphology presents one single peak (with two faint tails) strongly suggests that the gas heating was caused by the energetic collision of the two subclusters and that the passage of one of the subclusters through the other led to the formation of an X-ray wake \citep{Menanteau_2012}. The hot intracluster medium of each individual subcluster could also have generated some X-ray emission, but the hydrodynamic simulations show that the X-ray morphology and luminosity require a strong collision. The fact that the bolometric X-ray luminosity of El Gordo seems to be correlated with radio relics $-$ which have no obvious connection with the individual galaxy clusters $-$ further supports the hypothesis that the X-ray emission of El Gordo is mainly generated from the interaction of its two subclusters \citep[section~4.4 of][]{Menanteau_2012}.

Very low $V_{\textrm{infall}}$ values are also not supported by the observed peculiar velocity ($\Delta V_{\textrm{pec}} = 476 \pm 242~\textrm{km~s}^{-1}$) and inclination angle \citep[$\theta \le 35^\circ$;][]{Karen_2015} as these imply $V_{\textrm{obs}} = (234/\sin 35^{\circ})~\textrm{km~s}^{-1} = 408~\textrm{km~s}^{-1}$ in the most conservative scenario. This means that $V_{\textrm{obs}}$ would already be of the same order as $V_{\textrm{infall}} \approx 450~\textrm{km~s}^{-1}$ in the \citet{Kim_2021} model, which is in contradiction with the results of hydrodynamical simulations as these imply $V_{\textrm{infall}} \approx 1.5 \, V_{\textrm{obs}}$ \citep{Molnar_2015}.


The results of \citet{Karen_2015} do not give a clear estimate of $V_{\textrm{infall}}$ as they consider $V_{\textrm{peri}}$ more relevant for their study. Still, $V_{\textrm{infall}} \approx 1200~\textrm{km~s}^{-1}$ can be inferred by comparison with other studies. For this $V_{\textrm{infall}}$ value, El Gordo would be in $\approx 3.5\sigma$ tension with $\Lambda$CDM (see Figure~\ref{Results_plot}). However, \citet{Kim_2021} identified several problems with their MCMC analysis, so their nominal results might not be very reliable. More importantly, \citet{Karen_2015} also excluded all models in which the collision velocity exceeds the escape velocity of El Gordo, which makes their velocity estimates unsuitable to test $\Lambda$CDM as they implicitly assume its validity.

The first hydrodynamical simulations of El Gordo \citep{Donnert_2014} found that with $V_{\textrm{infall}} = 2600~\textrm{km~s}^{-1}$, the total X-ray luminosity of El Gordo can be reproduced, but not its two-tailed morphology. According to our analysis, this infall velocity is incompatible with $\Lambda$CDM at $> 5\sigma$ confidence. \citet{Donnert_2014} assumed a mass ratio and initial distance quite similar to those adopted in our analysis, but they assumed a somewhat higher mass ($M_{200} = 2.71 \times 10^{15} \, M_{\odot}$). Their $V_{\textrm{infall}}$ value is quite consistent with the best-fit $V_{\textrm{infall}} = 2500~\textrm{km~s}^{-1}$ obtained in the more detailed simulations of \citet{Zhang_2015} for a similar mass ($M_{200} \approx 3 \times 10^{15} \, M_{\odot}$). Given that in the \citet{Zhang_2015} simulations the higher-mass Model B ($M_{200} \approx 3 \times 10^{15} \, M_{\odot}$) requires a lower infall velocity than the lower-mass Model A ($M_{200} \approx 2 \times 10^{15} \, M_{\odot}$), we find it unlikely that assuming a higher mass had caused \citet{Donnert_2014} to overestimate $V_{\textrm{infall}}$.

\citet{Molnar_2015} explored a wider parameter space in their hydrodynamical simulations of El Gordo. They found that for $V_{\textrm{infall}} = 2250~\textrm{km~s}^{-1}$, they could fairly accurately reproduce the X-ray morphology, the positions of the X-ray peak and mass peaks, and the observed relative radial velocity. Still, the mean temperature and total X-ray luminosity obtained with this model are lower than observed \citep{Zhang_2015}. \citet{Molnar_2015} used a similar mass and mass ratio to the ones used in our analysis, but they chose a lower initial separation (4.25~Mpc). At our initial distance (6.06~Mpc) $-$ and under the assumption of energy conservation $-$ their $V_{\textrm{infall}}$ corresponds to
\begin{eqnarray}
    V_{\textrm{infall}} ~&=&~ \biggl[\left( 2250~\textrm{km~s}^{-1} \right)^2 \nonumber \\
    ~&+&~ 2 \, GM \left(\frac{1}{4.25~\textrm{Mpc}} - \frac{1}{6.06~\textrm{Mpc}}\right)\biggr] ^{1/2}\nonumber \\
    ~&=&~ 2522.29~\textrm{km~s}^{-1} \, ,
\end{eqnarray}
where $G$ is the gravitational constant and $M = 2.15 \times 10^{15} \, M_{\odot}$ is the total mass \citep{Molnar_2015}. This value corresponds to an $\approx 5.1\sigma$ tension with the $\Lambda$CDM model in our analysis ($\approx 5.5\sigma$ with the Bullet Cluster). This tension is slightly reduced by considering their other valid models with $V_{\textrm{infall}} = 2000~\textrm{km~s}^{-1}$ at 4.25~Mpc (2302.05~km~s$^{-1}$ at 6.06~Mpc), but with this velocity it is not possible to reproduce the X-ray brightness profile of El Gordo as accurately as with $V_{\textrm{infall}} = 2250~\textrm{km~s}^{-1}$.

\citet{Zhang_2015} presented two models of the El Gordo interaction in their hydrodynamical simulations. We focus on their Model A since this is the one that most closely resembles the latest observations and the assumptions made in our analysis. It has an infall velocity of $V_{\textrm{infall}} = 3000~\textrm{km~s}^{-1}$, which our analysis shows to be in more than $5\sigma$ tension with $\Lambda$CDM. Even though it still has some deficiencies (e.g. more asymmetric tidal tail, inclination angle is too large), this model is the only one so far that manages to reach the high temperature and luminosity of El Gordo, reproduce its surface brightness profile and mass density distribution and, to some extent, generate the two-tailed X-ray morphology, all while assuming a total mass of $\approx 2 \times 10^{15} \, M_{\odot}$ and a mass ratio of $\approx 2$, consistent with the latest weak-lensing results \citep{Kim_2021}.

The previous estimates of the El Gordo (or the joint El Gordo $+$ Bullet Cluster) tension with the $\Lambda$CDM cosmology were obtained accounting for the area of the survey regions in which El Gordo and the Bullet Cluster were found. Our more conservative estimates (dotted lines in Figure~\ref{Results_plot}) assume these survey regions covered the full sky, which would imply that El Gordo (and the Bullet Cluster) are the only problematic objects for $\Lambda$CDM that have been observed so far. However, other interacting galaxy clusters with extreme properties have already been found, including A1758 at $z = 0.279$ with two subclusters of total mass $M_{200} = 3.3^{+0.4}_{-0.6} \times 10^{15} \, M_{\odot}$ \citep{Schellenberger_2019} and MACS J1149.5+2223 at $z = 0.544$ with three subclusters of total mass $M_{500} = (1.04 \pm 0.05) \times 10^{15} \, M_{\odot}$ \citep{Bruno_2021}, Moreover, full-sky surveys such as the Planck survey have already identified several other clusters with very high masses at $z>0.3$, including PLCK G214.6+37.0 \citep[composed of three subclusters at $z \approx 0.47$ with total mass $M_{200} \approx 1.17 \times 10^{15} \, M_{\odot}$;][]{Planck_VIII_2011} and PLCK G287.0+32.9 \citep[a quadruple system of clusters at $z = 0.39$ with total mass $M_{200} = 2.04^{+0.20}_{-0.21} \times 10^{15} \, M_{\odot}$;][]{Planck_VI_2013}.\footnote{Note that at $z \approx 0.5$ and $z \approx 0.3$ structures with $M_{178} \gtrsim 1.7 \times 10^{15} \, M_{\odot}$ and $M_{178} \gtrsim 2.1 \times 10^{15} \, M_{\odot}$, respectively, are expected to be extremely rare in a $\Lambda$CDM cosmology \citep{Watson_2014}.} We therefore consider it reasonable to scale the number of candidate pairs to the survey volume in our nominal analysis.

However, we cannot be certain about whether there were any selection effects involved in the discovery of El Gordo and the Bullet Cluster. An important issue is whether wider less sensitive surveys had been used to get hints that extreme galaxy cluster collisions would be found in certain regions of the sky. If this were the case, the tension obtained in our nominal analysis might have been overestimated. The maximum impact that this could have on our results is given by our conservative scenario (dotted lines in Figure~\ref{Results_plot}), where we obtained the tension at each $V_{\textrm{infall}}$ for the case in which El Gordo (and the Bullet Cluster) are assumed to be the only objects in the sky contributing to the tension with $\Lambda$CDM. These results show that even in this extremely conservative scenario, the tension with $\Lambda$CDM is still $\approx 4.5\sigma$ for $V_{\textrm{infall}} \approx 2500~\textrm{km~s}^{-1}$ \citep[][Model B]{Molnar_2015, Zhang_2015}. This increases to $\approx 4.9\sigma$ for $V_{\textrm{infall}} = 3000~\textrm{km~s}^{-1}$, the infall velocity of Model A \citep{Zhang_2015}.

\section{Conclusions}
\label{Conclusions}

For the new weak-lensing mass estimate of El Gordo \citep{Kim_2021} and for all infall velocities consistent with the hydrodynamical simulations described in Section~\ref{Introduction}, El Gordo still falsifies $\Lambda$CDM at close to $5\sigma$. According to the nominal model of \citet{Molnar_2015}, infall velocities of $\approx 2500~\textrm{km~s}^{-1}$ are required to reproduce the morphological properties of El Gordo ($\approx 5.1\sigma$ tension). $V_{\textrm{infall}}$ needs to be higher in order to reproduce El Gordo's high temperature and luminosity: $V_{\textrm{infall}} \approx 2600~\textrm{km~s}^{-1}$ according to \citet{Donnert_2014}, which corresponds to $\approx 5.2\sigma$ tension. The only nominal models that manage to simultaneously reproduce both its morphology and its high luminosity and temperature are fiducial Model A and fiducial Model B from \citet{Zhang_2015}, with $V_{\textrm{infall}} = 3000~\textrm{km~s}^{-1}$ and $V_{\textrm{infall}} = 2500~\textrm{km~s}^{-1}$, respectively. Out of these two models, only Model A has a mass and mass ratio similar to the observed ones. Its $V_{\textrm{infall}}$ value corresponds to $\approx 5.4\sigma$ tension. The tension with $\Lambda$CDM is even higher ($\approx 5.7\sigma$) when we take into account the Bullet Cluster as another problematic object for $\Lambda$CDM. El Gordo and the Bullet Cluster are not the only observed galaxy clusters whose properties are highly unusual in a $\Lambda$CDM context. Because of this, we took into consideration the sky area in which El Gordo and the Bullet Cluster were found when obtaining our previous results. However, it is possible that the surveys might have already been targeting these particular objects based on previous wider surveys, which could have enabled their identification in a smaller survey region. To estimate the maximum extent to which this could have caused us to overestimate the tension with $\Lambda$CDM, we obtained the tension assuming the discovery surveys covered the full sky. We found that even in this very conservative scenario, the tension between these galaxy cluster collisions and $\Lambda$CDM amounts to $4.5\sigma-4.9\sigma$. We therefore conclude that El Gordo and the Bullet Cluster would still pose a significant challenge to $\Lambda$CDM even if selection effects had an influence on the survey areas in which they were discovered.


Given the recent more accurate results on El Gordo's mass and other properties, we expect that our results will encourage the scientific community to revisit the hydrodynamical simulations of El Gordo. This will clarify if a lower $V_{\textrm{infall}}$ compatible with $\Lambda$CDM can be consistent with the observed properties of El Gordo, or if this massive early collision does indeed point to a failure of the $\Lambda$CDM model.

\section*{Data availability}

The data underlying this article are available in the article.

\begin{acknowledgments}

E.A. is supported by a stipend from the Stellar Populations and Dynamics Research Group at the University of Bonn. I.B. is supported by Science and Technology Facilities Council grant ST/V000861/1. P.K. acknowledges support through the Deutscher Akademischer Austauschdienst-Eastern European Exchange Programme. The authors would like to thank the referee for comments and suggestions that have helped to improve this manuscript.
\end{acknowledgments}

\bibliographystyle{aasjournal}
\bibliography{El_Gordo_ApJ}
\end{document}